# Demonstration of a New Transport Regime of Photon in Two-dimensional Photonic Crystal

**Xiangdong Zhang**
Department of Physics, Beijing Normal University, Beijing 100875, China

## Abstract

A new transport regime of photon in two-dimensional photonic crystal near the Dirac point has been demonstrated by exact numerical simulation. In this regime, the conductance of photon is inversely proportional to the thickness of sample, which can be described by Dirac equation very well. Both of bulk and surface disorders always reduce the transmission, which is in contrast to the previous theoretical prediction that they increase the conductance of electron at the Dirac point of grephene. However, regular tuning of interface structures can cause the improvement of photon conductance. Furthermore, large conductance fluctuations of photon have also been observed, which is similar to the case of electron in graphene.

Recently, there has been a great deal of interest in studying the physical properties of graphene due to the successful fabrication experimented by Novoselov et al. [1]. Graphene is a monolayer of carbon atoms densely packed in a honeycomb lattice, which can be viewed as either an individual atomic plane pulled out of bulk graphite or unrolled single-wall carbon nanotubes. In graphene, the energy bands can be described at low energy by a two-dimensional (2D) Dirac equation centered on hexagonal corners (Dirac points) of the honeycomb lattice Brillouin zone [2, 3]. The quasiparticle excitations around the Dirac point obey linear Dirac-like energy dispersion. The presence of such Dirac-like quasiparticles leads to a new "pseudo-diffusive" transport regime for charge carriers in graphene [4,5]. Some unusual properties in the transport regime such as large conductance fluctuations and increase of conductance by disorder have also been pointed out theoretically [7-10]. However, these phenomena have not been demonstrated experimentally, because such a test is hindered by the difficulty to maintain a homogeneous electron density throughout the system [11].

Analogous to above electron system, the optical transmission near the Dirac point in 2D photonic crystal (PC) has also been discussed theoretically [12, 13]. In some 2D PCs with triangular or honeycomb lattices, the band gap may become vanishingly small at corners of the Brillouin zone, where two bands touch as a pair of cones. Such a conical singularity is also referred to as the Dirac point similar to the case of graphene. According to the analyses of Ref. [12, 13], photon transport near the Dirac point in the PC can also be described by the following Dirac equation

$$-iv_D(\nabla \cdot \vec{\sigma})\psi = (\omega - \omega_D)\psi \ . \qquad (1)$$

Where $\psi = (\psi_1, \psi_2)$ represents the amplitudes of two degenerate Bloch states at one of the corners of the hexagonal first Brillouin zone , and $\sigma = (\sigma_1, \sigma_2)$ is Pauli matrices. The frequency $\omega_D$ and velocity $v_D$ in the Dirac point depend on the structure of PC. For a PC slab with thickness $L$ in air background, we can obtain the dimensionless photon conductance $G \propto 1/L$ by applying the resolution of Dirac equation and boundary condition of flux conservation [13]. That is to say, the transmission of photon through 2D PC slab is inversely proportional to its thickness, which is similar to the case of electron transport in graphene. If such a theoretical result can be demonstrated successfully by the experimental measurements or exact numerical simulation, some unusual effects at the Dirac point such as large conductance fluctuation and the effect of disorder on conductance can also be tested, because the PC could provide an ideal testing ground, which some difficulties such as homogeneous electron density in grephene do not exist in it.

Based on such a research background, in this paper we investigate the transport properties of photon in the 2D PC near the Dirac point by using multiple-scattering method. The multiple-scattering method is a very efficient way of handling the scattering problem of a finite sample containing cylinders of circular cross sections, and it is capable of reproducing accurately the experimental transmission data, which should be regarded as exact numerical simulation [14].

We consider a 2D triangular lattice of cylinders immersed in air background with lattice constant $a$ . The radius ( $R$ ) and the dielectric constant ( $\varepsilon$ ) of the cylinders are $0.3a$ and 11.4, respectively. The band structure of the system for P wave obtained by the multiple-scattering Korringa–Kohn–Rostoker method is shown in Fig. 1(a). The corresponding dimensionless photon conductance $G$ (the ratio between the total transmitted flux and incident photon flux) along $\Gamma K$ direction for the sample with $L = 9a$ and $18a$ has also been plotted in Fig.1(b) as solid line and dotted line, respectively. The excellent agreement between the conductance and the band structure can be observed. The key feature of this band structure is the presence of a pair of Dirac points at $f = 0.466\ \omega a / 2\pi c$ . In the following, we calculate the sample size dependences of conductance at the frequencies near or away from such a Dirac point as marked by red arrows in Fig.1(b).

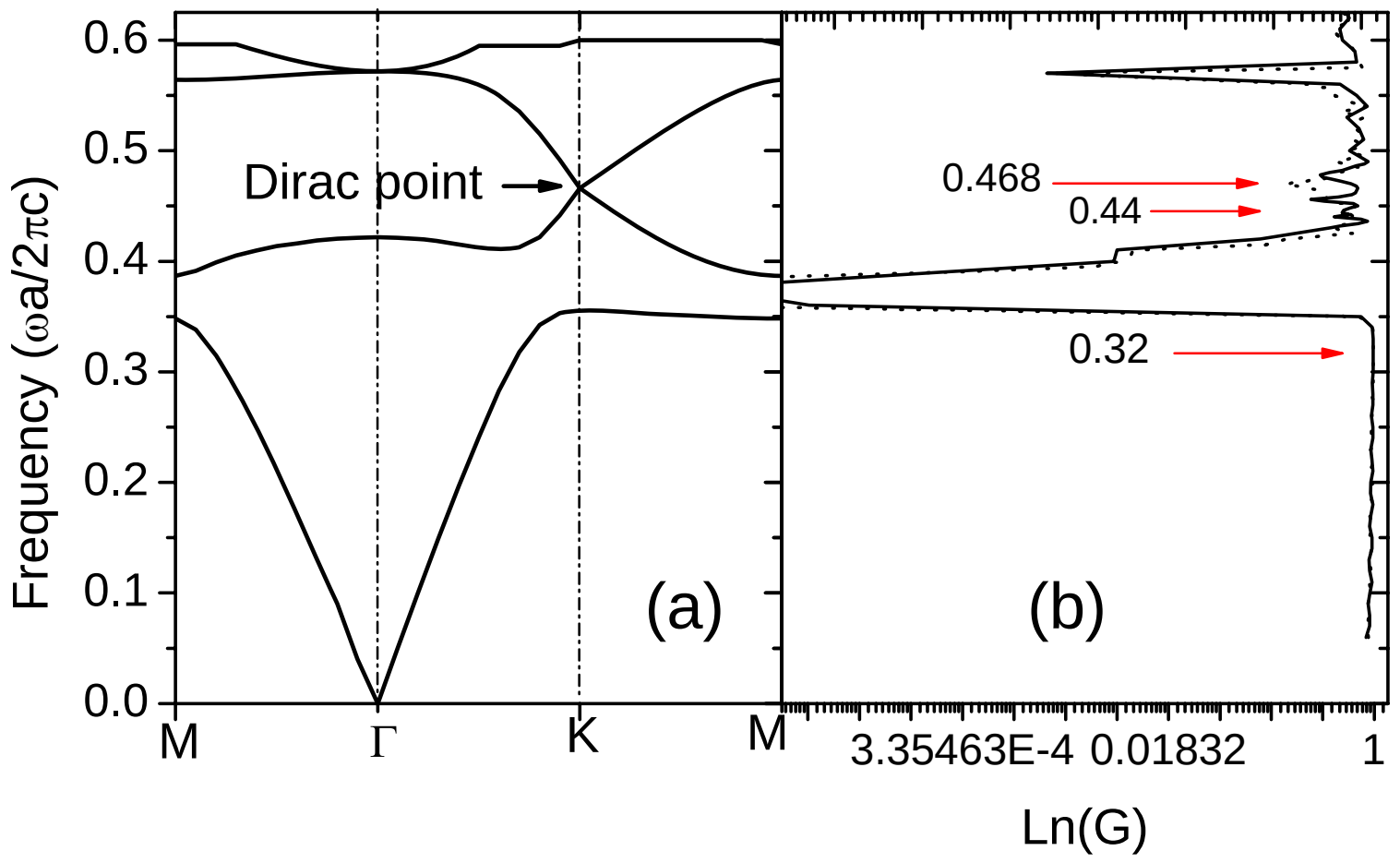

**Fig. 1 (a) Calculated photonic band structure of P wave for a triangular lattice of dielectric cylinder with R/a=0.3 and $\varepsilon = 11.4$ in an air background. (b) The conductance of photon for 2D PC corresponding to (a). The solid line corresponds to the result with the sample thickness $L = 9a$ and dotted line to that with $L = 18a$ .**

Thus, we take slab samples with width $W$ and thickness $L$, denoted as $W \times L$. The source is prepared by passing a plane wave through an open slit in front of the sample. The width of the slit is about 20% smaller than the sample width to avoid diffraction. In all the calculations discussed below, $W$ is chosen to be $80\,a$. The product of $G$ along $\Gamma K$ direction and $L$ as a function of the sample thickness at $f = 0.468\ \omega a / 2\pi c$ (near the Dirac point) is plotted in Fig.2 as triangular dot. For compare, we also give the results at $f = 0.44\ \omega a / 2\pi c$ (away from the Dirac point ) and $f = 0.32\ \omega a / 2\pi c$ (another band) in Fig.2 as circle dot and square dot, respectively. It is seen clearly that they exhibit different feature although all these frequencies are in band region. At $f = 0.468\ \omega a / 2\pi c$, $G \times L$ keeps constant (red line) with the increase of $L$ although it oscillates around red line. Here the red line is drawn only for view. The oscillating characteristic depends on the interface and the thickness of sample, but they leave the red line unaffected. This indicates that the conductance is inversely proportional to the thickness of sample, which is similar to diffusion behavior of wave through a disordered medium even in the absence of any disorder in the PC. This is completely different from the case at $f = 0.32\ \omega a / 2\pi c$. The conductance is near constant and $G \times L$ increase linearly with the increase of $L$, which exhibits the standard ballistic behavior characteristic. The change feature of conductance at $f = 0.44\ \omega a / 2\pi c$ is at the middle between above two kinds of case. It oscillates around green line (drawn only for view) and the average values increase with the increase of $L$, but slope of green line is smaller than that of square-dot line ($f = 0.32\ \omega a / 2\pi c$). This means that the conductance possesses both of ballistic and diffusion behaviors in this case.

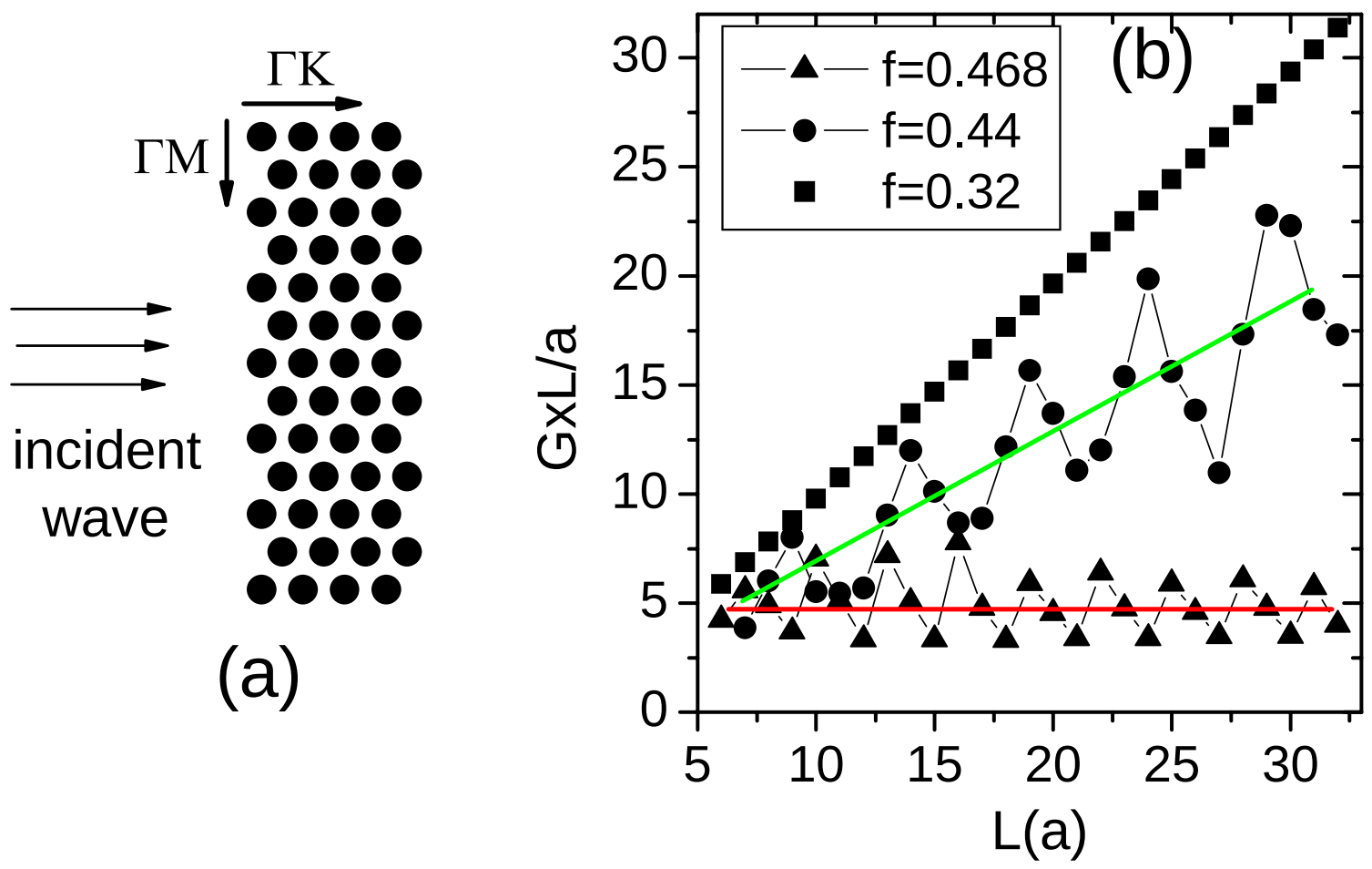

**Fig.2 (a) Schematic picture depicting simulation processes. (b) The product of $G$ and $L$ as a function of sample thickness at different frequencies. The frequencies are in unit of $\omega a / 2\pi c$ . Red line and green line are plotted only for view. The other parameters are identical to those in Fig.1.**

From the above calculated results, we find that the transmission properties of photon near the Dirac point are different from the ballistic behavior in general band region. At the same time, it is also different from the cases of exponential decay with the increase of sample thickness when the frequency lies in the gap region. Thus, it can be regarded as a new transport regime, which is similar to the properties of electron transport at the Dirac point of grephene. Such a pseudo-diffusive feature can be described by Dirac equation very well [12, 13].

After the new transport regime of photon in the PC has been demonstrated, we test the effect of disorder on the conductance in the following. In general, there are three typical kinds of randomness in the PC structures: cylinder site displacements (site randomness), cylinder radius variations (size randomness), and varied dielectric constants of cylinders (dielectric randomness). We first consider the case of site randomness. Such a disordered PC can be produced by randomizing the above ordered system. The degree of randomness can be controlled. Each cylinder is moved randomly, but only within a range $[-dr, dr]$, where $dr < (a - 2R)/2$ . For various configurations with different $dr$ , the conductance and its variance $VarG = \langle G^2 \rangle - \langle G \rangle^2$ are calculated exactly by the multiple-scattering method. Every process is repeated for 500 configurations. Averages of the conductance as a function of the strength of disorder at different frequencies are shown in Fig.3. Circle dots and triangular dot represent the results with $L = 10a$ and $20a$ , respectively. We find that the conductance always decrease with the increase of the random strength for different frequencies and sample thicknesses. The effect of disorder on the conductance near the Dirac point is smaller than those at other frequencies. This is different from the previous theoretical prediction that the disorders increase the conductance of electron at the Dirac point of grephene.

In contrast to monotonous change of the conductance, the fluctuation exhibits different feature with the increase of disorder. Figure 4 (a) and (b) show the variance $(Var(G))$ of photon conductance as a function of the random strength at different frequencies for $L = 10a$ and $20a$ , respectively. Square dots, circle dots and triangular dots represent the results at $f = 0.32, 0.44$ and 0.468 $\omega a / 2\pi c$ , respectively. It is shown clearly that the peaks appear at certain random strengths for $f = 0.44$ and 0.468 $\omega a / 2\pi c$ . This can be understood from comparing the results in Fig.3 with those in Fig.4(c) and (d).

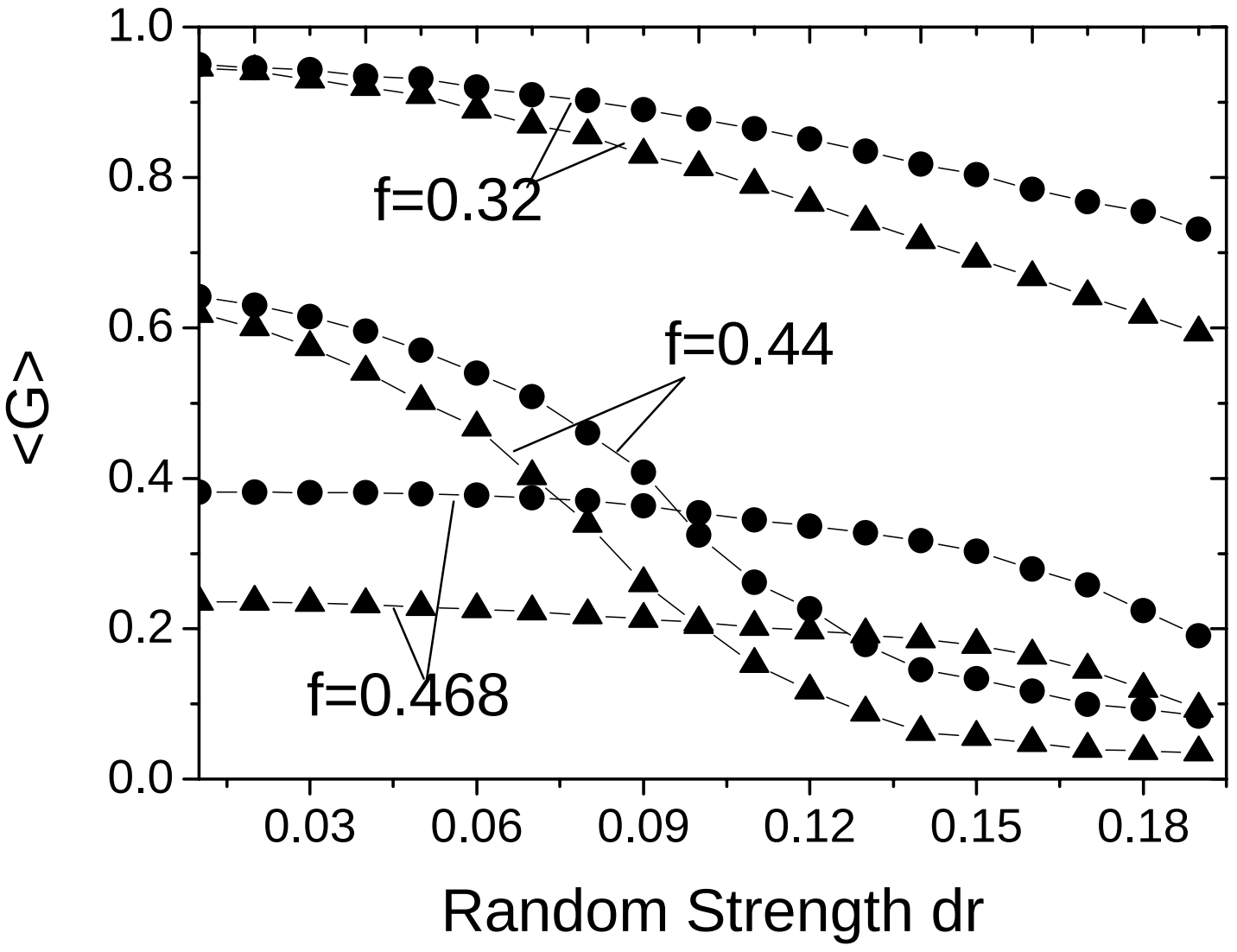

**Fig.3 Averages of the conductance as a function of the random strength for different thickness of the sample at various frequencies. Circle dots are for $L = 10a$ and triangular dots for $L = 20a$. The frequencies are marked in figure in unit of $\omega a / 2\pi c$. The other parameters are identical to those in Fig.1.**

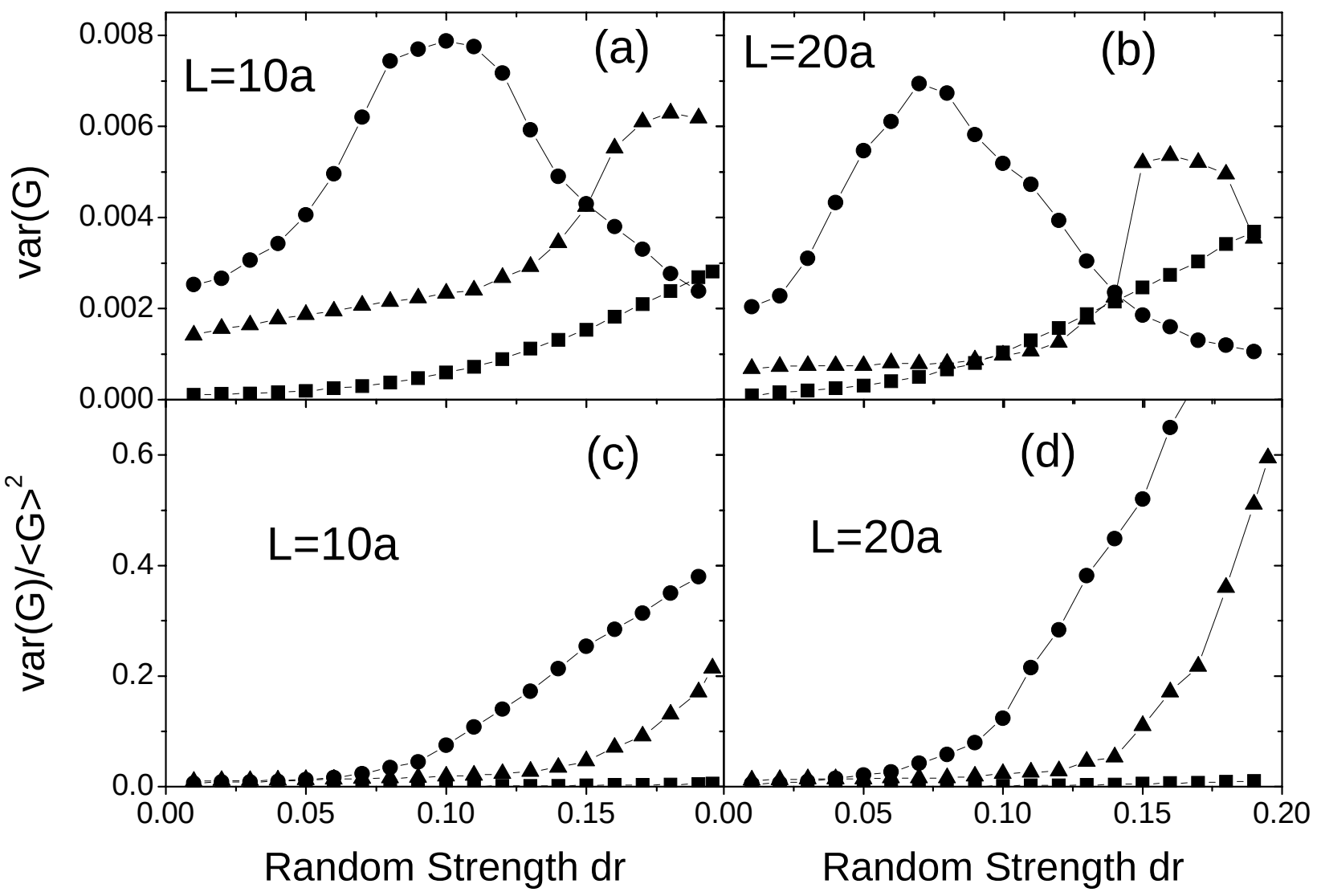

**Fig.4 Variances ((a) and (b)) and relative fluctuations ((c) and (d)) of the conductance as a function of the random strength for the sample with different thicknesses at various frequencies. Square dots, circle dots and triangular dots correspond to $f = 0.32, 0.44$ and $0.468\ \omega a / 2\pi c$, respectively. The thicknesses of the sample are marked in figures. The other parameters are identical to those in Fig.1.**

Fig.4 (c) and (d) describe relative fluctuation of conductance ($Var(G) / < G >^2$) as a function of the random strength at different frequencies for $L = 10a$ and $20a$, respectively. The curves in Fig.4(c) and (d) correspond completely to those in Fig.4(a) and (b). We notice that the change features of relative fluctuation at different frequencies are similar. All of them increase monotonously with the increase of random strength. However, their values and change magnitudes with the increase of random strength are very different. At $f = 0.44$ and 0.468 $\omega a / 2\pi c$ , the relative fluctuations are large and change rapidly with the increase of random strength. On one hand, the relative fluctuation increases rapidly with the increase of random strength. On the other hand, the conductance decreases with the increase of random strength. Integrating two aspects leads to the appearance of peaks for the lines with the circle dots and the triangular dots in Fig.4(a) and (b). These calculated results for photon are similar to the previous observation of electron in graphene [7]. An analytical theory to explain such a phenomenon keeps to be developed.

The above discussions only focus on the case of bulk disorder. In fact, one is also interesting in the case of interface disorder because it inevitably occurs in the fabrication of the sample. In previous discussions on electron transport at the Dirac point of grephene, some theories have predicted that the conductance of electron can be improved by the interface disorder [7-9]. In the following, we test such a case for photon. The dielectric constants of cylinders on two surface layers of PC slab are taken randomly within a range $[\varepsilon - \Delta\varepsilon, \varepsilon + \Delta\varepsilon]$. This is the case of dielectric randomness. The calculated results of average conductance as a function of $\Delta\varepsilon$ at $f = 0.32, 0.44$ and 0.468 $\omega a / 2\pi c$ for $L = 14a$ are plotted in Fig.5(a) as square dots, circle dots and triangular dots, respectively. We see clearly that the change features of conductance with surface disorder is similar to the case of bulk disorder, they all decrease with the increase of random strength. We do not find the case of increase. In contrast, if we change the dielectric constant of the surface layers regularly, the conductance can be improved at appropriate cases. Dotted line and dashed line in Fig.5(b) represent such cases. Solid line, dashed line and dotted line in Fig.5(b) describe the photon conductance as a function of dielectric constants of cylinders on the surface layers at $f = 0.32, 0.44$ and 0.468 $\omega a / 2\pi c$ , respectively. For the case with only ballistic characteristic (solid line), the conductance does not be affected by the surface structures. However, once the conductance possesses diffusion feature partly or completely (dashed line or dotted line), they will be affected largely. They are oscillating functions of dielectric constants of cylinders on the surface layers. The conductance is actually improved at appropriate surface structures. The root of the phenomenon lies in the excitation of surface modes and the appearance of resonant transmissions for some certain surface structures, which is identical with the previous analysis in Ref.[15].

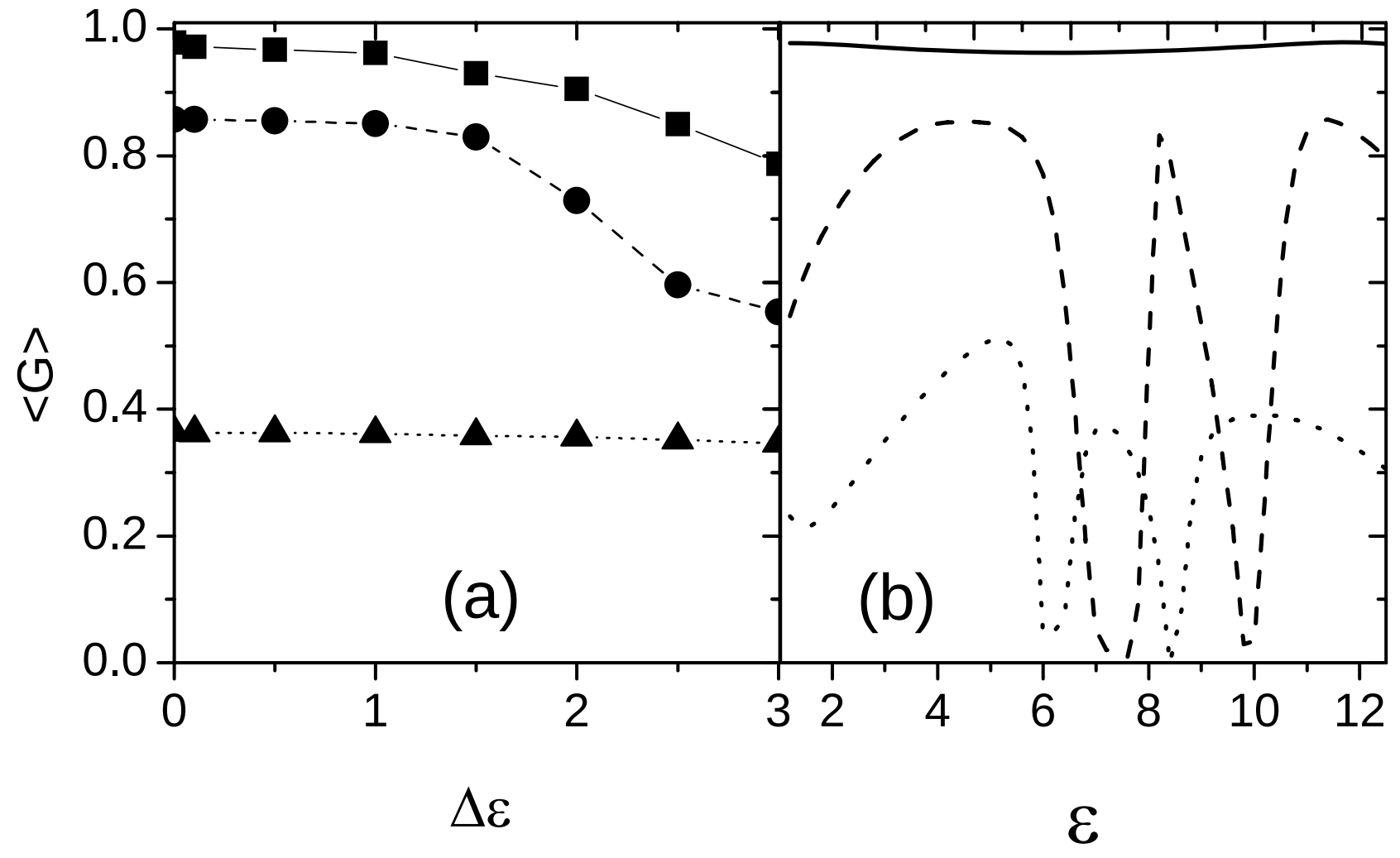

**Fig.5 (a) Averages of the conductance as a function of the random dielectric constants of cylinders on the surface layers for the sample with $L = 14a$. Square dots, circle dots and triangular dots correspond to the results at $f = 0.32, 0.44$ and $0.468\ \omega a / 2\pi c$, respectively. (b) Photon conductance as a function of dielectric constants of cylinders on the surface layers. Solid line, dashed line and dotted line correspond to the results at $f = 0.32, 0.44$ and $0.468\ \omega a / 2\pi c$, respectively. The other parameters are identical to those in Fig.1.**

In summary, we have presented an exact numerical calculation of photon conductance in the 2D PC near or away from the Dirac point. We have demonstrated existence of a new transport regime of photon in the PC near the Dirac point. In this regime, the conductance of photon is inversely proportional to the thickness of sample, which can be described by Dirac equation very well. This is completely different from the cases of exponential decay with the increase of sample thickness when the frequency lies in the gap and ballistic behavior in general band region. Both of bulk and surface disorders always reduce the transmission, which is different from the previous theoretical prediction that they increase the conductance of electron at the Dirac point of grephene. In contrast, the conductance of photon can be improved by tuning regularly interface structures due to the excitation of surface modes and resonant transmission. Furthermore, large conductance fluctuations of photon have also been observed, which is similar to the case of electron in graphene.

This work was supported by the National Natural Science Foundation of China (Grant No. 10674017). The project was also sponsored by NCET and RFDP.